\documentclass[journal,twoside,web]{ieeecolor}

\usepackage{generic}
\usepackage{cite}
\usepackage{url}
\usepackage{hyperref}
\usepackage[pdftex]{graphicx}
\usepackage{multirow}
\usepackage{amsmath}
\usepackage{mathrsfs}
\usepackage{amsfonts,amssymb}
\usepackage{booktabs}
\usepackage{algpseudocode}
\usepackage{algorithm}
\usepackage{makecell}

\algnewcommand{\IfThenElse}[3]{
  \State \algorithmicif\ #1\ \algorithmicthen\ #2\ \algorithmicelse\ #3}

\def\BibTeX{{\rm B\kern-.05em{\sc i\kern-.025em b}\kern-.08em
    T\kern-.1667em\lower.7ex\hbox{E}\kern-.125emX}}
\begin{document}
\title{Lung-DDPM: Semantic Layout-guided Diffusion Models for Thoracic CT Image Synthesis}
\author{Yifan Jiang, Yannick Lemaréchal, Sophie Plante, Josée Bafaro, Jessica Abi-Rjeile, Philippe Joubert, Philippe Després and Venkata Manem
\thanks{(Corresponding author: Yifan Jiang and Venkata Manem.) Y. Jiang, and V. Manem are with Centre de recherche du CHU de Québec-Université Laval, 2705 Bd Laurier, Québec, QC G1V 4G2, Canada (e-mail: yfjiang.works@hotmail.com; venkata.manem@crchudequebec.ulaval.ca). Y. Lemaréchal, S. Plante, J. Bafaro, J. Abi-Rjeile, P. Joubert, P. Després are with Centre de recherche de l’Institut universitaire de cardiologie et de pneumologie de Québec-Université Laval, Québec, Canada. }}

\maketitle

\begin{abstract}
With the rapid development of artificial intelligence (AI), AI-assisted medical imaging analysis demonstrates remarkable performance in early lung cancer screening. However, the costly annotation process and privacy concerns limit the construction of large-scale medical datasets, hampering the further application of AI in healthcare. To address the data scarcity in lung cancer screening, we propose Lung-DDPM, a thoracic CT image synthesis approach that effectively generates high-fidelity 3D synthetic CT images, which prove helpful in downstream lung nodule segmentation tasks. Our method is based on semantic layout-guided denoising diffusion probabilistic models (DDPM), enabling anatomically reasonable, seamless, and consistent sample generation even from incomplete semantic layouts. Our results suggest that the proposed method outperforms other state-of-the-art (SOTA) generative models in image quality evaluation and downstream lung nodule segmentation tasks. Specifically, Lung-DDPM achieved superior performance on our large validation cohort, with a Fréchet inception distance (FID) of 0.0047, maximum mean discrepancy (MMD) of 0.0070, and mean squared error (MSE) of 0.0024. These results were 7.4×, 3.1×, and 29.5× better than the second-best competitors, respectively. Furthermore, the lung nodule segmentation model, trained on a dataset combining real and Lung-DDPM-generated synthetic samples, attained a Dice Coefficient (Dice) of 0.3914 and sensitivity of 0.4393. This represents 8.8\% and 18.6\% improvements in Dice and sensitivity compared to the model trained solely on real samples. The experimental results highlight Lung-DDPM’s potential for a broader range of medical imaging applications, such as general tumor segmentation, cancer survival estimation, and risk prediction. The code and pretrained models are available at \url{https://github.com/Manem-Lab/Lung-DDPM/}.
\end{abstract}


\section{Introduction}
\label{sec1}
\IEEEPARstart{L}{ung} cancer is recognized as the most lethal cancer type globally, accounting for the highest mortality rates in both genders. GLOBOCAN 2022 reports 2.48 million new cases and 1.82 million deaths worldwide in 2022 \cite{bray2024global}. In the United States alone, the estimated lung cancer-related cases and deaths for 2024 are 234,580 and 125,070, respectively \cite{siegel2024cancer}. The 5-year survival rate of lung cancer highly depends on the diagnostic stage. Data from patients diagnosed with lung cancer in England between 2016 and 2020 show that 5-year survival rates for early-stage diagnoses are relatively high (Stage I: 65\%, Stage II: 40\%), while they drop significantly for late-stage diagnoses (Stage III: 15\%, Stage IV: 5\%) \cite{mohamed2024can}. The low survival rate in late-stage patients highlights the urgent clinical need for early lung cancer detection.

Leveraging the strong capability in feature representation and high generalization of AI across various data types, it is now widely utilized in different computer-aided diagnosis (CAD) tasks, including medical image segmentation \cite{huang2024segment}, cancer risk prediction \cite{mikhael2023sybil}, and survival prediction \cite{cui2022survival}. Modern AI technology heavily relies on high-quality datasets, which require massive efforts in collection and annotation. In medical imaging, due to costly annotation processes by professionals and patient privacy regulations, acquiring well-labeled datasets with an adequate number of patients is increasingly challenging. In lung cancer screening, it is common for only lung nodules or tumors to be annotated, while the annotation of other thoracic anatomical structures is often deemed impractical. According to Ma et al. \cite{ma2021toward}, annotating lung regions and lesions in COVID-19 thoracic CT scans takes approximately 400 minutes per scan. Additionally, patient privacy regulations like Health Insurance Portability and Accountability Act (HIPAA) \cite{annas2003hipaa} and the General Data Protection Regulation (GDPR) \cite{beaulieu2019privacy} prevent researchers from accessing patient data. Therefore, costly annotation processes and patient privacy regulations lead to low-quality and insufficient datasets. Oakden-Rayner \cite{oakden2020exploring} analyzed two large-scale X-ray datasets and found that inaccurate labels lead to performance deterioration in deep learning models. Cho et al. \cite{cho2015much} emphasized that adequate training samples in multiple medical imaging analysis tasks are essential for promising performance. Therefore, rapidly advancing deep learning techniques strongly demand solutions to data scarcity in medical imaging.

\begin{figure*}[!ht]
\centering
\includegraphics[width=16cm]{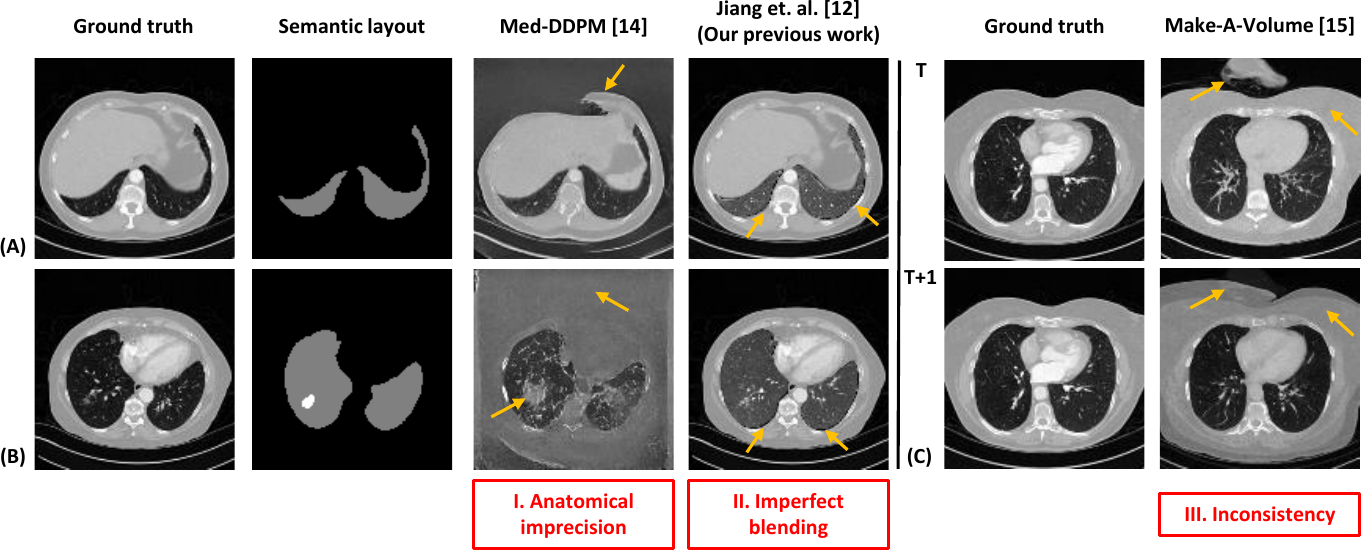}
\caption{Demonstration of existing issues of SOTA generative models in thoracic CT image synthesis. In subfigures (A) and (B), anatomical imprecision and imperfect blending issues are presented. The first column is the reference CT images (ground truth), and the second column is the input semantic layouts (white: lung nodule, gray: lung, black: background), third to fifth columns show the existing anatomical imprecision and imperfect blending issues using synthetic images from Med-DDPM \cite{dorjsembe2024conditional} and our previous work \cite{jiang2020covid}. In subfigure (C), we illustrate the inconsistency issue. The first column is the adjunct reference CT images at the T and T+1 slice positions, respectively. The second column is the corresponding synthetic images from Make-A-Volume \cite{zhu2023make}.  Yellow arrows are indicators of interesting points.}
\label{fig1} 
\end{figure*}

To address data scarcity in medical imaging, researchers have recently favored deep generative models to synthesize artificial data for dataset augmentation. Generative adversarial networks (GANs) and diffusion models constitute the backbone of medical image synthesis models. Our previous work \cite{jiang2020covid} introduced a conditional GANs-based method for COVID-19 CT image synthesis. Leveraging dual generator \& discriminator and dynamic communication mechanisms, the network generates high-quality 2D COVID-19 CT images.  HA-GAN \cite{sun2022hierarchical} introduced a novel end-to-end GANs architecture for generating high-resolution 3D medical images, using a hierarchical structure during training to manage memory constraints and ensure anatomical consistency while enabling full high-resolution image generation during inference. More recently, diffusion model-based methods have gained significant attention due to their superior performance in image generation tasks. Med-DDPM \cite{dorjsembe2024conditional} proposed a diffusion model for 3D semantic brain MRI synthesis that addresses data scarcity and privacy concerns through semantic conditioning, producing high-quality, diverse images that outperform existing methods showing potential for data augmentation and image anonymization in medical imaging. Make-A-Volume \cite{zhu2023make} introduced a novel diffusion-based framework for cross-modality 3D medical image synthesis, leveraging a 2D latent diffusion backbone with added volumetric layers to achieve high-quality, volumetrically consistent 3D image generation while maintaining computational efficiency. GEM-3D \cite{zhu2024generative}, as the successor of Make-A-Volume, further decomposes 3D medical images into semantic layouts and patient-specific informed slices containing anatomical appearance, position, and scanning pattern information. This enables one-to-many mapping from semantic layouts to volumes and improves image quality. 

\begin{figure*}[!ht]
\centering
\includegraphics[width=16cm]{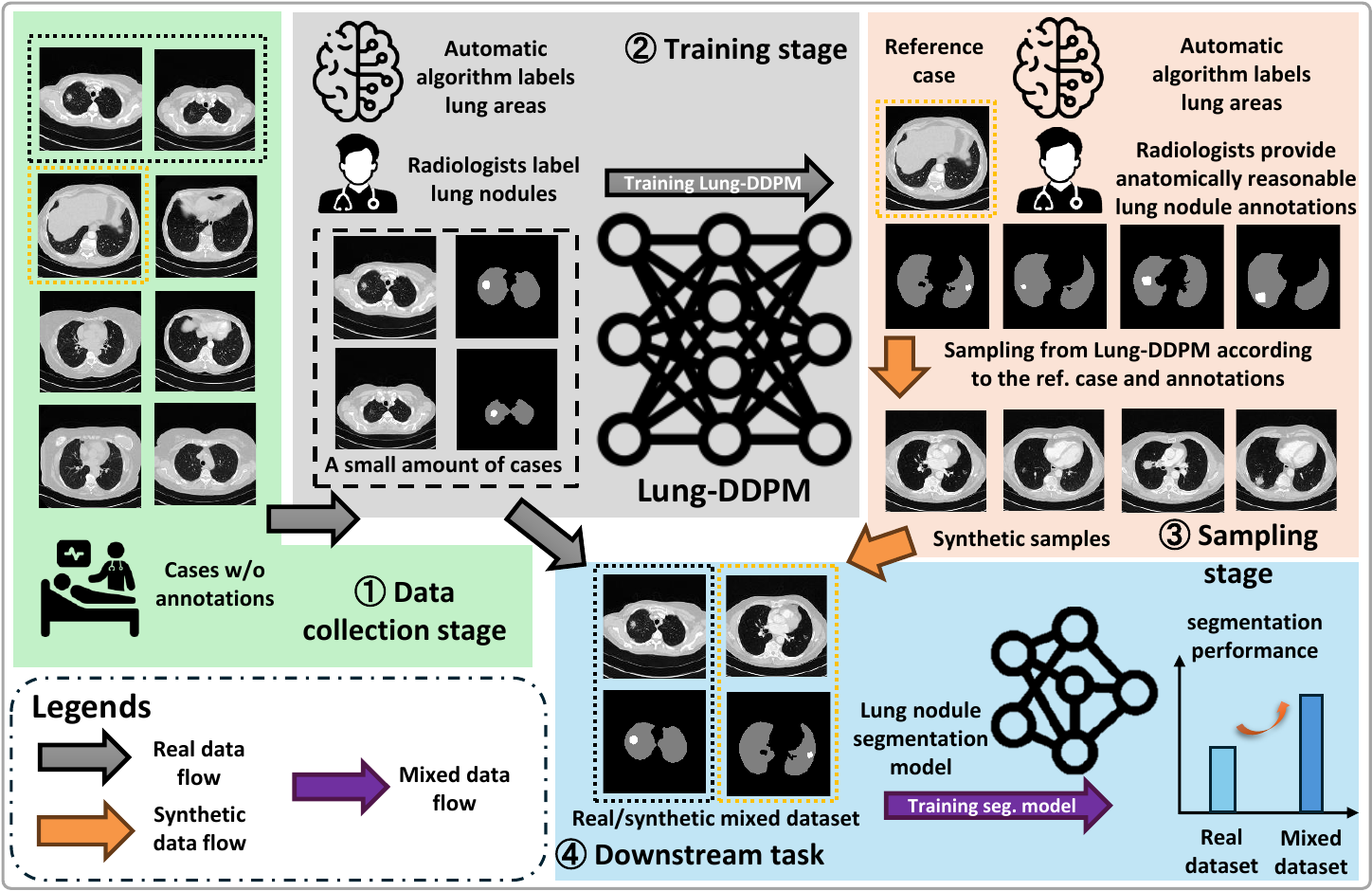}
\caption{The workflow of the proposed method can be divided into four main stages: \textcircled{\raisebox{-0.9pt}{1}} Data collection stage (green area), \textcircled{\raisebox{-0.9pt}{2}} Training stage (grey area), \textcircled{\raisebox{-0.9pt}{3}} Sampling stage (pink area), and \textcircled{\raisebox{-0.9pt}{4}} Downstream task (blue area). There are three different data flows: real data (grey arrow), synthetic data (orange arrow), and mixed data (purple arrow). The black dashed box indicates the real data involved in stages \textcircled{\raisebox{-0.9pt}{2}} and \textcircled{\raisebox{-0.9pt}{4}}. The orange dashed box indicates the synthetic data derived from the real data in stages \textcircled{\raisebox{-0.9pt}{3}} and \textcircled{\raisebox{-0.9pt}{4}}.}
\label{fig2} 
\end{figure*}

Although above methods can address medical image synthesis tasks in certain modalities, these SOTA methods have three major issues that urgently need to be addressed: \textbf{I. Anatomical imprecision; II. Imperfect blending; III. Inconsistency}. In Figure \ref{fig1} (A) and (B), we observe that Med-DDPM suffers from generating anatomically imprecise CT images in extra-pulmonary areas, which also impacts image quality in lung areas. This anatomical imprecision is caused by the lack of semantic layout guidance for extra-pulmonary areas. Nevertheless, acquiring complete semantic layouts for thoracic CT scans in practical scenarios is too expensive and time-consuming, creating a circular problem. To mitigate the anatomical imprecision issue, our previous work \cite{jiang2020covid} attempted to generate CT images only in lung areas and simply blend them with corresponding images of extra-pulmonary areas. However, this simple blending strategy introduced a new imperfect blending issue, causing joint seams and tone discrepancy between lung and extra-pulmonary areas. As technology advances, emerging lightweight generative models allow researchers to synthesize 3D CT volumes with lower computational resources, while introducing inconsistency issues between adjacent slices. As depicted in \ref{fig1} (C), this inconsistency issue usually occurs in memory-efficient network designs \cite{zhu2023make}\cite{zhu2024generative} that break a whole generation task into parts and construct a complete sample using synthetically partial samples.

To address the aforementioned issues and better alleviate medical data scarcity in lung cancer screening, we introduce \textbf{Lung-DDPM}, a denoising diffusion probabilistic models-based thoracic CT image synthesis approach. Specifically, we designed an anatomically aware sampling process, which guides the sampling process for the diffusion model jointly using a reference CT volume and the corresponding lung semantic layout. This anatomically aware sampling process enables the proposed method to dynamically blend synthetic lung images and extra-pulmonary images during the sampling procedure, synthesizing high-quality thoracic CT images with reasonable anatomical structures. We summarize the main contributions of this paper as follows:

\begin{itemize}
\item [(1)] To address the anatomical imprecision issue, we introduce a novel anatomically aware sampling (AAS) process that jointly utilizes reference CT volumes and corresponding lung annotations to guide the denoising process and dynamically blend synthetic lung areas with reference extra-pulmonary areas during sampling.
\item [(2)] To mitigate the imperfect blending issue, we present an adjusted   3D U-Net with residual and attention blocks, extending our previous work to 3D CT volume synthesis. This network cooperates with the proposed AAS process to generate seamless synthetic samples.
\item [(3)] To tackle the inconsistency issue, we develop a diffusion model-based thoracic CT image synthesis approach that accepts and generates whole thoracic CT volumes, avoiding inconsistencies caused by partial generation.
\item [(4)] Our proposed method outperforms other SOTA generative models in image quality evaluation and downstream lung nodule segmentation tasks, demonstrating its advanced performance in various applications.
\item [(5)] This thoracic CT image synthesis method significantly reduces the workload of medical staff in annotation tasks. In practice, radiologists need only provide reasonable lung nodule semantic layouts. The proposed method can then generate an unlimited number of synthetic CT scans by referencing a single real CT scan. These synthetic CT scans are valuable in alleviating data scarcity for AI-assisted lung cancer screening.
\end{itemize}

\section{Proposed method}
\label{sec2}
\subsection{Workflow}
In this section, we detail the proposed diffusion model-based thoracic image synthesis method. The overall workflow is illustrated in Figure \ref{fig2} and comprises four stages:
\begin{itemize}
\item [\textcircled{\raisebox{-0.9pt}{1}}] \textbf{Data collection stage:} The workflow begins with the collection of thoracic CT scans from clinical practice to establish a case management database. At this stage, no annotations are applied to the cases.
\item [\textcircled{\raisebox{-0.9pt}{2}}] \textbf{Training stage:} The collected data is annotated by experienced radiologists (for lung nodules) and automatic algorithms (for lung areas) to balance accuracy and efficiency. The resulting annotated dataset is then used to train the proposed Lung-DDPM model.
\item [\textcircled{\raisebox{-0.9pt}{3}}] \textbf{Sampling stage:} Reference cases are randomly selected from the dataset and annotated for lung areas using an automatic algorithm. Radiologists then provide anatomically reasonable lung nodule annotations for these reference cases. The pretrained model from stage \textcircled{\raisebox{-0.9pt}{2}} is subsequently used to generate synthetic samples conditioned on the reference cases and their annotations.
\item [\textcircled{\raisebox{-0.9pt}{4}}] \textbf{Downstream task stage:} In this stage, the synthetic samples are combined with real data to create a real/synthetic mixed dataset, which is used to train a lung nodule segmentation model. Training with this mixed dataset is expected to improve performance compared to training with real data alone.
\end{itemize}

\begin{figure*}[!ht]
\centering
\includegraphics[width=16cm]{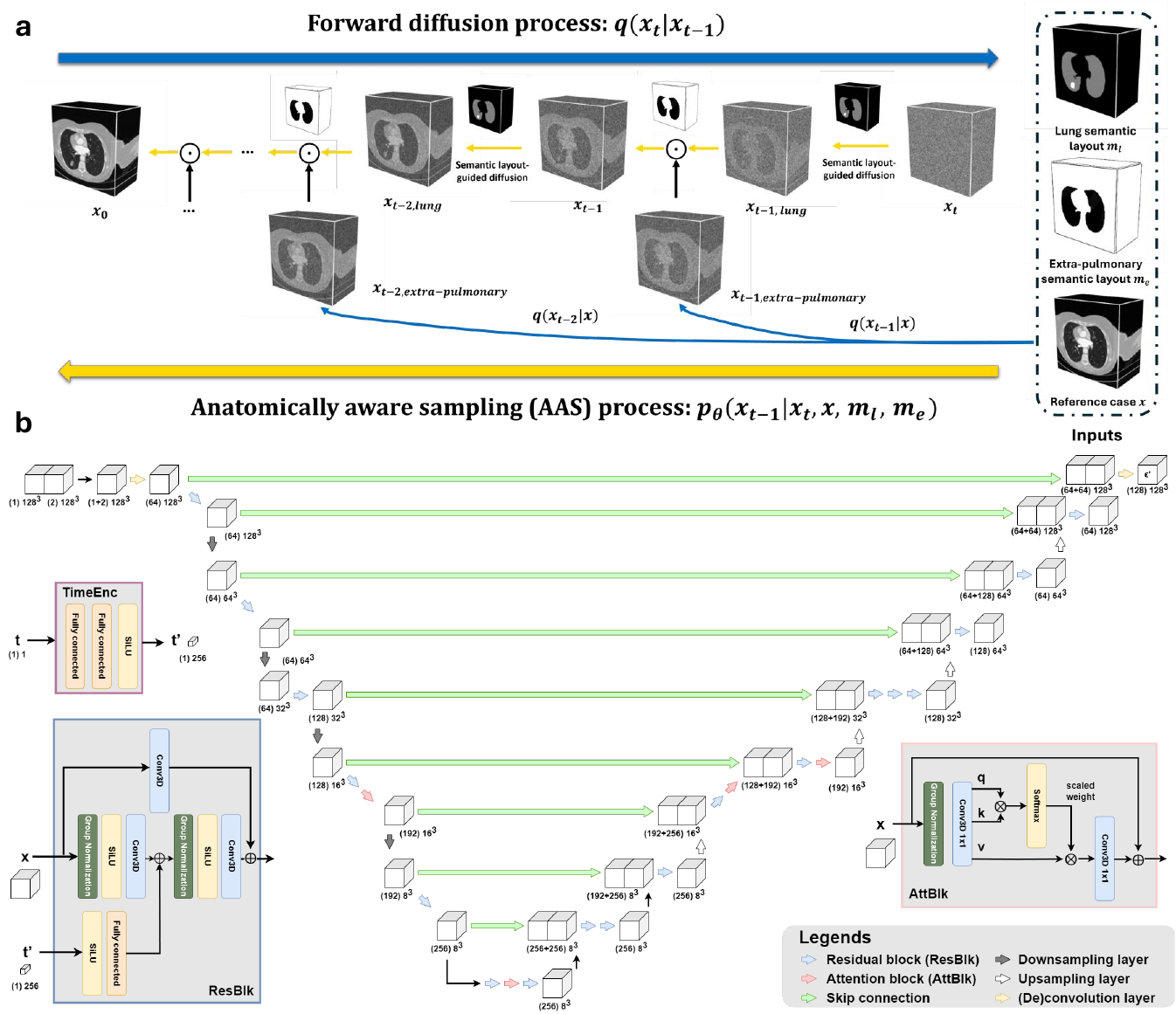}
\caption{Details of the proposed method. \textbf{a}, The proposed semantic layout-guided diffusion model incorporates an anatomically aware sampling (AAS) process, which consists of two components: a forward diffusion process (blue arrow) and an AAS process (yellow arrow). \textbf{b}, The adapted 3D U-Net structure is illustrated. Residual blocks are indicated by blue arrows and box, attention blocks by pink arrows and box, downsampling layers by grey arrows, upsampling layers by white arrows, skip connections by green arrows, (de)convolution layers by yellow arrows, and time step encoders (TimeEnc) by a purple box. The numbers below each cube represent the intermediate feature size in the format '(channel number) side length3'.}
\label{fig3} 
\end{figure*}

\subsection{Training and sampling processes}
Figure \ref{fig3}a illustrates our proposed semantic layout-guided diffusion model with an anatomically aware sampling (AAS) process. This process consists of two components: a forward diffusion process and an AAS process.

\noindent \textbf{Forward diffusion process}: Given a 3D CT volume $x_0 \sim q(x_0)$, a forward diffusion process gradually adds Gaussian noise to the input $x_0$, generating a series of intermediate latents $x_1, \ldots, x_{t-1}, x_t$ with a variance $\beta_t \in (0,1)$:

\begin{equation}
\begin{aligned}
q(x_t|x_{t-1}) = \mathcal{N}(\sqrt{1-\beta_t}x_{t-1}, \beta_t\mathbf{I})
\end{aligned}
\label{eq1}
\end{equation}

Any $x_t$ at time step $t$ can be directly sampled from $x_0$:

\begin{equation}
\begin{aligned}
x_t=\sqrt{\bar{\alpha_t}}x_0+\sqrt{1-\bar{\alpha_t}}\epsilon
\end{aligned}
\label{eq2}
\end{equation}

where $\epsilon \sim \mathcal{N}(0,\mathbf{I})$, $\alpha_t=1-\beta_t$, and $\bar{\alpha_t}=\prod_{k=0}^{t}{\alpha_k}$. \\
\\
\noindent \textbf{Anatomically aware sampling (AAS) process}: In a traditional sampling process \cite{ho2020denoising,dorjsembe2024conditional}, a neural network $\theta$ acts as a noise predictor, estimating the noise $\epsilon_{\theta} (x_t, t)$. The denoising process $x_{t-1} \sim p_\theta (x_{t-1}|x_t)$ can then be formulated as follows:

\begin{equation}
\begin{aligned}
x_{t-1} = \frac{1}{\sqrt{\alpha_t}} \left(x_t - \frac{1 - \alpha_t}{\sqrt{1 -\bar{\alpha_t}}} \epsilon_{\theta} (x_t\oplus m, t)\right) + \sigma_t z
\end{aligned}
\label{eq3}
\end{equation}
where $m$ is the semantic layout, $z \sim \mathcal{N}(0,\mathbf{I})$, $\sigma_t=\sqrt{\beta_t}$, and $\oplus$ is the concatenating operation between the semantic layout $m$ and the intermediate latent $x_t$. However, the traditional sampling process has a significant drawback: its overdependence on detailed semantic layouts. As demonstrated in Figure \ref{fig1}, without guidance from semantic layouts for extra-pulmonary areas, this process may produce abnormal results with incorrect anatomical structures. In clinical practice, manually annotating every anatomical structure in a thoracic CT scan is impractical. Moreover, semi-automatic and automatic segmentation tools \cite{ma2024segment,isensee2021nnu} are not sufficiently reliable to provide consistent semantic layouts due to the diverse types and parameter settings of CT scanners \cite{weisman2023multi}.

To address this issue, we propose a novel anatomically aware sampling process. This process leverages only radiologist-provided lung semantic layouts $m_l$, extra-pulmonary semantic layouts $m_e$ (which can be easily derived from the corresponding $m_l$), and a reference case $x$ to achieve better image quality and more precise anatomical structures for synthetic CT images. For any intermediate latent $x_t$, we first acquire $x_{t-1,lung}$ through a semantic layout-guided diffusion process:
\begin{equation}
\begin{aligned}
x_{t-1,lung} = \frac{1}{\sqrt{\alpha_t}} \left(x_t - \frac{1 - \alpha_t}{\sqrt{1 -\bar{\alpha_t}}} \epsilon_{\theta} (x_t\oplus m_l, t)\right) + \sigma_t z
\end{aligned}
\label{eq4}
\end{equation}

Next, we sample $x_{t-1,extra-pulmonary}$ through the forward diffusion process, starting from a reference case $x$:
\begin{equation}
\begin{aligned}
x_{t-1,extra-pulmonary} = \sqrt{\bar{\alpha}_{t-1}}x+\sqrt{1-\bar{\alpha}_{t-1}}\epsilon
\end{aligned}
\label{eq5}
\end{equation}

To obtain $x_{t-1}$, we blend $x_{t-1,lung}$ and $x_{t-1,extra-pulmonary}$ using the extra-pulmonary semantic layout $m_e$:
\begin{equation}
\begin{aligned}
x_{t-1} = (1-m_e) \otimes x_{t-1,lung} + m_e \otimes x_{t-1,extra-pulmonary}
\end{aligned}
\label{eq6}
\end{equation}
where $m_e=1-m_l$ serves to spatially separate lung regions from extra-pulmonary regions during the sampling process. Note that even though $m_e$ itself does not contain anatomical semantic details, it acts as a spatial mask that enables dynamic fusion between synthesized lung regions (guided by $m_l$) and extra-pulmonary regions sampled from a real reference CT image. Through the proposed AAS process, we can generate high-quality thoracic CT images guided by radiologist-provided semantic layouts while maintaining anatomically reasonable extra-pulmonary areas, which significantly impact overall image quality. The training and AAS pseudo-codes are depicted in Algorithms \ref{alg1} and \ref{alg2}, respectively.

\begin{algorithm}
\caption{Training procedure}
\textbf{Input:} real CT image $x_0\sim q(x_0)$ and the corresponding lung semantic layout $m_l$\\
\textbf{Output:} Trained noise predictor $\hat{\epsilon}_{\theta}$
\begin{algorithmic}[1]
\Repeat 
\State $t\sim Uniform(\{ 1,...T \})$
\State $\epsilon \sim \mathcal{N}(0,\textbf{I})$
\State $x_t=\sqrt{\bar{\alpha_t}}x_0+\sqrt{1-\bar{\alpha_t}}\epsilon$
\State $\hat{x}_t=x_t\oplus m_l$
\State Taking a gradient step towards
\State \hskip1.5em $\nabla_\theta \mathcal{L}(\epsilon,\epsilon_\theta (\hat{x}_t, t))$
\Until{converged}
\end{algorithmic}
\label{alg1}
\end{algorithm}

\begin{algorithm}
\caption{Anatomically aware sampling (AAS) procedure: given a pretrained noise predictor $\hat{\epsilon}_{\theta}$.}
\textbf{Input:} reference case $x$, the corresponding lung semantic layout $m_l$ and extra-pulmonary semantic layout $m_e$\\
\textbf{Output:} synthetic thoracic CT images $x_0$ according to $x$, $m_l$ and $m_e$
\begin{algorithmic}[1]
\State sample $x_T\sim \mathcal{N}(0, \textbf{I})$
\For{\textbf{all} $t$ from $T$ to 1}
    \State $\epsilon'=\hat{\epsilon}_{\theta}(x_t\oplus m_l,t)$
    \IfThenElse {$t > 1$}
      {$z\sim \mathcal{N}(0, I)$}
      {$z = 0$}
    \State $x_{t-1,lung} = \frac{1}{\sqrt{\alpha_t}} (x_t - \frac{1 - \alpha_t}{\sqrt{1 -\bar{\alpha_t}}} \epsilon') + \sigma_t z$
    \State $x_{t-1,extra-pulmonary} = \sqrt{\bar{\alpha}_{t-1}}x+\sqrt{1-\bar{\alpha}_{t-1}}\epsilon$
    \State $x_{t-1} = (1-m_e)\otimes x_{t-1,lung} + m_e \otimes x_{t-1,extra-pulmonary}$
\EndFor
\State \Return $x_0$
\end{algorithmic}
\label{alg2}
\end{algorithm}

\subsection{Network designs}
We adapted the 3D U-Net $\epsilon_\theta$ based on the network designs of 3D-DDPM \cite{dorjsembe2022three} and Med-DDPM \cite{dorjsembe2024conditional} to integrate with the proposed AAS process. The adapted 3D U-Net network design is shown in Figure \ref{fig3}b.

\noindent \textbf{3D U-Net}: We build upon the well-established U-Net backbone used in the original DDPM framework \cite{ho2020denoising}, which integrates residual blocks (ResBlks) and attention blocks (AttBlks) to enhance 3D representation learning. In our work, we extend this design to volumetric medical imaging by adapting it for 3D thoracic CT synthesis. While the use of ResBlks and AttBlks aligns with prior architectures, their integration within a 3D denoising diffusion framework introduces unique challenges in spatial modeling and memory efficiency, which we address through careful architectural calibration. Specifically, we modified it to accept a noise matrix and the corresponding lung semantic layout, both with dimensions of $128^3$. The network input consists of a noised CT image and its corresponding semantic layouts. After passing through a convolution layer, the encoder path (the left part of the 3D U-Net) progressively reduces spatial dimensions while increasing the number of feature channels, enabling the network to learn hierarchical representations of thoracic CT images. The bottleneck (the middle part of the 3D U-Net) processes low-dimensional features to capture global context and relationships within the CT image representations. Finally, the decoder (the right part of the 3D U-Net) reconstructs the spatial resolution of the features, gradually transforming the compact representation back into a full-sized output. Simultaneously, it combines both low-level and high-level features from the encoder via skip connections to generate detailed CT images. The decoder output is the predicted noise $\epsilon'$ for the input image.

\noindent \textbf{Residual block (ResBlk)}: In Figure \ref{fig3}b, we illustrate the design of the ResBlk. It takes the intermediate feature x and the embedded time step $t'$ from the time step encoder as inputs. The embedded time step $t'$ enables the block to adjust its behavior according to the current denoising stage. The latent representations are then passed through both convolution layers and residual connections, ensuring that unmodified features can flow through the network, preventing gradient vanishing in deeper layers. Notably, as the resolution of CT images increases, the ResBlk becomes more critical in this 3D U-Net structure.

\noindent \textbf{Attention block (AttBlk)}: This block integrates channel attention mechanisms into the 3D U-Net, allowing each part of the CT image to influence every other part—crucial for maintaining coherence as noise is removed. Additionally, it enables the 3D U-Net to capture global anatomical structures that are challenging for convolutional layers alone to model. We illustrate the AttBlk in Figure \ref{fig3}b. The AttBlk takes the intermediate features $x$ as input and transforms them into three components: $q$ (Query), $k$ (Key), and $v$ (Value) through separate $1\times 1$ convolutional layers. The dot product of $q$ and $k$ is computed, measuring how closely each key matches the query. The result is scaled, and a softmax function is applied to obtain attention weights, which are used to compute a weighted sum of the values.

Utilizing the adapted 3D U-Net with the modified ResBlk and AttBlk, we can predict the noise $\epsilon'$ from $x_t$ for each time step $t$, and subsequently obtain $x_{t-1,lung}$.

\subsection{Learning object}
According to Med-DDPM \cite{dorjsembe2024conditional}, although both L1 and L2 losses are commonly used in DDPM literature, L2 loss (unnormalized MSE loss) is more sensitive to outliers and tends to produce noisier results. In our model design, we follow the findings of Med-DDPM and adopt L1 loss as the learning objective to measure the differences between the added noise $\epsilon$ and the predicted noise $\epsilon'$:

\begin{equation}
\begin{aligned}
\mathcal{L}(\theta)=\mathbb{E}_{x_0\sim q(x_0),\epsilon\sim\mathcal{N}(0,\textbf{I}),t}[|\epsilon-\epsilon'|]
\end{aligned}
\label{eq7}
\end{equation}

\section{Experiments}
\label{sec3}
\subsection{Description of cohorts}
We utilize an internal (Quebec Heart and Lung Institute Research Center (IUCPQ)) dataset and an external dataset (LIDC-IDRI \cite{samuel2011lung}) in this study. The population characteristics of the IUCPQ dataset are summarized in the online supplementary material. Additionally, we provide the patient list from the LIDC-IDRI dataset used in this study on the project's GitHub page.

\begin{table*}[htbp] 
 \centering
 \caption{Organization of our datasets (Synth. indicates synthetic samples.)} 
 \begin{tabular}{c c c c} 
  \toprule 
  Cohort & size & Usage stage & Data type \\ 
  \midrule 
 Discovery (upstream)  & 800 & Model training (Lung-DDPM) & Real \\
 Validation (IUCPQ) & 800 & Synthetic sample generation & Real $\rightarrow$ Synthetic \\ 
 Validation (LIDC-IDRI) & 366 & External synthetic sample generation & Real $\rightarrow$ Synthetic \\ 
 \midrule 
 Discovery (downstream)  & 800$\sim$1600 & Segmentation model training (Real + Synthetic) & Mixed (Real + Synth) \\
 Test  & 249 & Segmentation performance evaluation & Real  \\
  \bottomrule 
 \end{tabular} 
\label{table2}
\end{table*}

\noindent \textbf{IUCPQ\footnote{The project was approved by the Comité d'éthique de la recherche of the Institut universitaire de cardiologie et de pneumologie de Québec-Université Laval (IUCPQ-UL), protocol number 2022-3752, 22156, on November 16, 2021, under the direction of Dr. Philippe Joubert.}:}  The internal dataset includes 1,849 patients who underwent lung cancer resection and were collected by the IUCPQ. Each case has an original CT image and the corresponding lung/lung nodule semantic layouts. Every case in the dataset is nodule-positive. Lung nodules were annotated by Dr. Philippe Joubert and Josée Bafaro. To reduce the workload of the annotation team, we acquired lung semantic layouts by applying a U-Net-based automatic segmentation algorithm [19]. We split the data into four cohorts, each serving a specific role (Table \ref{table2}).

\begin{table*}[htbp] 
\footnotesize
 \centering
 \caption{Image quality evaluation results. (The best evaluation score is marked in bold. $\uparrow$ means higher number is better, and $\downarrow$ indicates lower number is better. Seg. indicates semantic layouts, and ref. indicates reference frames.)} 
 \begin{tabular}{c c c c c c c c} 
  \toprule 
  Validation cohort & & \multicolumn{3}{c}{IUCPQ} & \multicolumn{3}{c}{LIDC-IDRI}\\ 
  \midrule 
  Method & Condition & FID ($\downarrow$) & MMD ($\downarrow$) & MSE ($\downarrow$) & FID ($\downarrow$) & MMD ($\downarrow$) & MSE ($\downarrow$)\\ 
  \midrule 
  HA-GAN \cite{sun2022hierarchical} & N/A & 0.0403±0.0002 & 0.1063±0.0002 & - & 0.0510±0.0000  & 0.1277±0.0000 & -\\
  Pix2pix3D \cite{wang2018high} &  Seg.  & 0.0346±0.0300 & 0.0217±0.0000 & 0.0712±0.0000 & 0.0129±0.0002 & 0.0497±0.0000 & 0.0581±0.0000 \\
  Med-DDPM \cite{dorjsembe2024conditional} & Seg. & 0.0464±0.0328 & 0.6806±0.0118 & 0.0707±0.0011 & 0.0371±0.0015 & 0.4039±0.0144 & 0.0767±0.0006 \\
 MAV \cite{zhu2023make} & Seg. & 0.1497±0.0134 & 1.4442±0.0074 & 0.1143±0.0001 & 0.0652±0.0009 & 0.5463±0.0034 & 0.1072±0.0002 \\
 GEM-3D \cite{zhu2024generative} & Seg. \& Ref. & 0.0557±0.0140 & 0.6312±0.0349 & 0.1484±0.0014 & 0.0313±0.0063 & 0.2897±0.0222 & 0.0803±0.0020   \\
 \makecell{Lung-DDPM \\ (OURS)} & Seg. \& Ref. & \textbf{0.0047±0.0080} & \textbf{0.0070±0.0002} & \textbf{0.0024±0.0000} & \textbf{0.0083±0.0077} & \textbf{0.0240±0.0011} & \textbf{0.0032±0.0000}\\
  \bottomrule 
 \end{tabular} 
\label{table3}
\end{table*}

\noindent \textbf{LIDC-IDRI:} To further validate the developed Lung-DDPM under public and transparent experimental conditions, we included the external LIDC-IDRI \cite{samuel2011lung} dataset, developed through collaboration between the Lung Image Database Consortium (LIDC) and the Image Database Resource Initiative (IDRI). This dataset consists of diagnostic and lung cancer screening thoracic CT scans with marked-up annotated lesions. It contains 1,010 cases with clinical thoracic CT scans, of which 875 have lung nodule annotations acquired through a two-phase image annotation process performed by four experienced thoracic radiologists. To obtain a single ground truth, a 50\% consensus criterion \cite{kubota2011segmentation} is adopted to manage the variability among different radiologists. Since existing automatic segmentation algorithms cannot stably annotate lung regions in contrast-enhanced CT scans, we excluded 509 cases with contrast-enhanced CT scans and constructed the final dataset using the remaining 366 cases. This cohort was used only for synthetic data generation and validation, not for training Lung-DDPM, as shown in Table \ref{table2}.

\subsection{Experimental settings}
To comprehensively evaluate the proposed method's performance and compare it to SOTA generative models, we designed an image quality evaluation and a lung nodule segmentation evaluation (downstream task). In this section, we will describe the experimental designs and results in detail.

\subsubsection{Evaluation metrics}
To accurately assess model performance and make it comparable among SOTA models, we utilize the following image quality metrics and medical imaging semantic segmentation metrics:

In our experiments, we consider three image quality metrics: Fréchet Inception Distance (FID)~\cite{heusel2017gans}, Maximum Mean Discrepancy (MMD)~\cite{gretton2012kernel}, and Mean Squared Error (MSE). FID measures the distance between the distributions of real and synthetic data, represented by deep learning features. We follow the implementation of HA-GAN~\cite{sun2022hierarchical} and utilize a 3D ResNet pretrained on radiological datasets~\cite{chen2019med3d} to extract features from CT scans. MMD compares real and synthetic data distributions by evaluating moment differences in a high-dimensional feature space using the same pretrained model. MSE is computed as $
\text{MSE}(x_{\text{real}}, x_{\text{syn}}, m_l) = \frac{1}{\sum_i m_l^{(i)}} \sum_{i=1}^{N} m_l^{(i) } \left(x_{\text{real}}^{(i)} - x_{\text{syn}}^{(i)}\right)^2$,
where \( x_{\text{real}} \) and \( x_{\text{syn}} \) are voxel intensities of the real and generated images, and \( m_l \) is the binary lung mask indicating pulmonary regions. This metric is used to assess structural consistency between paired real and synthetic images conditioned on the same semantic layout. Note that for the only unconditional generative model HA-GAN, we do not report MSE since there is no one-to-one correspondence between generated and real images, making pixel-wise comparison ill-defined.

There are three medical imaging semantic segmentation metrics that are considered in our experiments: Dice Coefficient (Dice), sensitivity, and specificity [29][30]. Dice is a widely used metric for assessing the overlap between the predicted segmentation and the ground truth. Sensitivity measures the proportion of actual positive voxels that were correctly identified; it indicates how well the segmentation model identifies all the voxels that belong to the structure of interest. Specificity, on the other hand, measures the proportion of actual negative voxels that were correctly identified as negative. It indicates how well the algorithm excludes voxels that don't belong to the structure of interest.

Moreover, we report the above evaluation metrics with a 95\% confidence interval calculated among 5 validation folds on the test cohort.

\subsubsection{Implementation details}
All CT images and their corresponding semantic layouts were preprocessed to a size of $128^3$. We trained the proposed model for 100,000 steps with a learning rate of 1e-5, using the Adam optimizer. We followed the same Exponential Moving Average (EMA) strategy as in \cite{dorjsembe2024conditional}, with a 0.995 decay factor. For the diffusion model parameters, we used a cosine noise schedule with 250 steps. The batch size was set to 1 during training. All experiments were conducted on the HPC of Université Laval using an Nvidia Tesla A100 80GB GPU.

\begin{figure}[!ht]
\centering
\includegraphics[width=8cm]{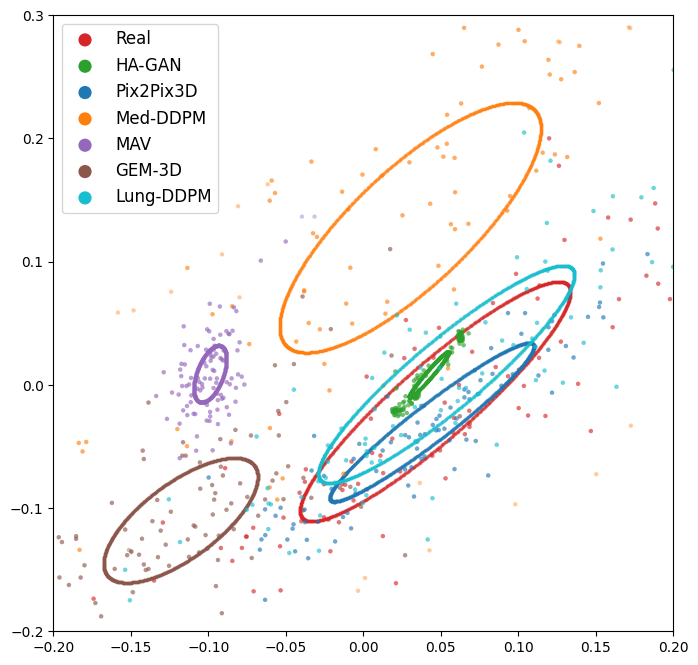}
\caption{The illustration of embedded synthetic samples using multidimensional scaling (MDS). We show real samples in red spots, synthetic samples from HA-GAN in green spots, Pix2pix3D in blue spots, Med-DDPM in orange spots, MAV in purple spots, GEM-3D in brown spots, Lung-DDPM in cyan spots.}
\label{fig4} 
\end{figure}

\begin{table*}[htbp] 
 \centering
 \caption{Experimental results for the lung nodule segmentation evaluation \hspace{\textwidth} \hspace{\textwidth} (The best and the second evaluation score are marked in bold and underline, respectively. $\uparrow$ means higher number is better, and $\downarrow$ indicates lower number is better. R. indicates real samples, and S. indicates synthetic samples.)} 
 \begin{tabular}{c c c c c} 
  \toprule 
  \makecell{Discovery cohort \\ (downstream)} & \makecell{Additional \\ data source} & Dice ($\uparrow$) & Sensitivity ($\uparrow$) & Specificity ($\uparrow$) \\ 
  \midrule 
  \makecell[cc]{800 R.: \\ 100\% discovery cohort (upstream)} & N/A & 0.3598±0.0976 & 0.3704±0.1094 & 0.9999±0.0000 \\
  \midrule 
  \makecell[cc]{800 R. + 183 R. or S.: \\ 100\% discovery cohort (upstream) + \\ 50\% validation cohort (LIDC-IDRI)}  & 
  \makecell{R. \\ MAV \cite{zhu2023make} \\ GEM-3D \cite{zhu2024generative} \\ Med-DDPM \cite{dorjsembe2024conditional} \\ Lung-DDPM (OURS)} & \makecell{\underline{0.3571±0.0969} \\ 0.3190±0.1594 \\ 0.2991±0.2047 \\ 0.3537±0.1003 \\ \textbf{0.3908±0.0452}} & 
  \makecell{0.3737±0.1535 \\ 0.3290±0.1688 \\ 0.3487±0.1398 \\ \underline{0.4046±0.2032} \\ \textbf{0.4393±0.1386}} & 
  \makecell{\underline{0.9999±0.0001} \\ \textbf{0.9999±0.0000} \\ 0.9980±0.0051 \\ 0.9998±0.0002 \\ \underline{0.9999±0.0001}}\\
 \midrule 
   \makecell[cc]{800 R. + 400 R. or S.: \\ 100\% discovery cohort (upstream) + \\ 50\% validation cohort (IUCPQ)}  & 
   \makecell{R. \\ MAV \cite{zhu2023make} \\ GEM-3D \cite{zhu2024generative} \\ Med-DDPM \cite{dorjsembe2024conditional} \\ Lung-DDPM (OURS)} & \makecell{\textbf{0.4044±0.0269} \\ 0.3196±0.2232 \\ 0.3852±0.0553 \\ 0.3804±0.0534 \\ \underline{0.3866±0.0293}} & 
   \makecell{\textbf{0.4053±0.0345} \\ 0.3380±0.2389 \\ \underline{0.4030±0.0509} \\ 0.3835±0.0695 \\ 0.3917±0.0449} & 
   \makecell{\textbf{0.9999±0.0000} \\ \underline{0.9999±0.0001} \\ \textbf{0.9999±0.0000}  \\ \textbf{0.9999±0.0000} \\ \textbf{0.9999±0.0000}}\\
 \midrule 
   \makecell[cc]{800 R. + 800 R. or S.: \\ 100\% discovery cohort (upstream) + \\ 100\% validation cohort (IUCPQ)}  & 
   \makecell{R. \\ MAV \cite{zhu2023make} \\ GEM-3D \cite{zhu2024generative} \\ Med-DDPM \cite{dorjsembe2024conditional} \\ Lung-DDPM (OURS)} & \makecell{\textbf{0.4178±0.0212} \\ 0.3577±0.0255 \\ 0.3584±0.0682 \\ 0.3638±0.0530 \\ \underline{0.3914±0.0282}} & 
   \makecell{\textbf{0.4377±0.0301} \\ 0.3674±0.0430 \\ 0.3891±0.0638 \\ 0.3729±0.0589 \\ \underline{0.4078±0.0519}} & 
   \makecell{\textbf{0.9999±0.0000} \\ \textbf{0.9999±0.0000} \\ \underline{0.9998±0.0003}  \\ \textbf{0.9999±0.0000} \\ \textbf{0.9999±0.0000}}\\
  \bottomrule 
 \end{tabular} 
\label{table4}
\end{table*}

\section{Results}\label{sec4}
\subsection{Quantitative results}
\subsubsection{Image quality evaluation}
In this evaluation, we focus on verifying the quality of synthetic images and comparing it with various SOTA generative models. First, we train models with the discovery (upstream) cohort and then sample 5-fold synthetic images according to the validation cohort, which we use to calculate image quality metrics. For IUCPQ, we generate 800 samples per fold (4000 samples in total), and for LIDC-IDRI, 366 samples per fold (1830 samples in total). We compare our method with HA-GAN \cite{sun2022hierarchical}, Pix2pix3D \cite{wang2018high}, Med-DDPM \cite{dorjsembe2024conditional}, Make-A-Volume (MAV) \cite{zhu2023make}, and GEM-3D \cite{zhu2024generative}. These SOTA methods can be categorized according to their conditions. The complete evaluation results are shown in Table \ref{table3}.

The proposed Lung-DDPM outperforms other competitors in all three evaluation metrics, with a FID of 0.0047, MMD of 0.0070, and MSE of 0.0024 on the IUCPQ validation cohort, and with a FID of 0.0083, maximum mean discrepancy MMD of 0.0240, and MSE of 0.0032 on the IUCPQ validation cohort. Compared to competitors that utilize no condition or a single condition (best performance metrics on the IUCPQ cohort, FID: 0.1497, MMD: 0.0217, and MSE: 0.0707. best performance metrics on the LIDC-IDRI cohort, FID: 0.0129, MMD: 0.0497, and MSE: 0.0581.), our method leverages the AAS process, integrating detailed semantic layouts and reference frames as a joint condition, which leads to superior performance in image quality evaluation. Unlike GEM-3D (achieves FID: 0.0557, MMD: 0.6312, and MSE: 0.1484 on the IUCPQ cohort. FID: 0.0313, MMD: 0.2897, and MSE: 0.0803 on the LIDC-IDRI cohort.), which uses incomplete reference frames and complete semantic layouts as conditions, our method utilizes complete reference frames and semantic layouts to ensure anatomically reasonable generation and stable synthetic sample quality. The image quality evaluation results indicate that the proposed method achieves performance that is an order of magnitude advanced compared to the SOTA competitors.

To evaluate synthetic samples' diversity and fidelity and illustrate the results intuitively, we follow \cite{sun2022hierarchical} and use the multidimensional scaling (MDS) method to visualize the embedded features of synthetic samples from competitors and our proposed method in Figure 4. Specifically, we first extract embedded features from 800 samples generated by competitors and our method according to the validation cohort (IUCPQ) using a pretrained 3D ResNet \cite{chen2019med3d}. We then project these features into two-dimensional space using MDS and fit them with an ellipse using least squares. A larger overlap between an ellipse and the red ellipse (representing real samples) indicates higher fidelity, while a larger absolute area of an ellipse represents higher diversity. As shown in Figure \ref{fig4},   Lung-DDPM shares the largest area with real samples while having an area similar to that of the real samples, implying that our proposed method achieves a better balance between diversity and fidelity compared to other competitors.

\begin{figure*}[!ht]
\centering
\includegraphics[width=16cm]{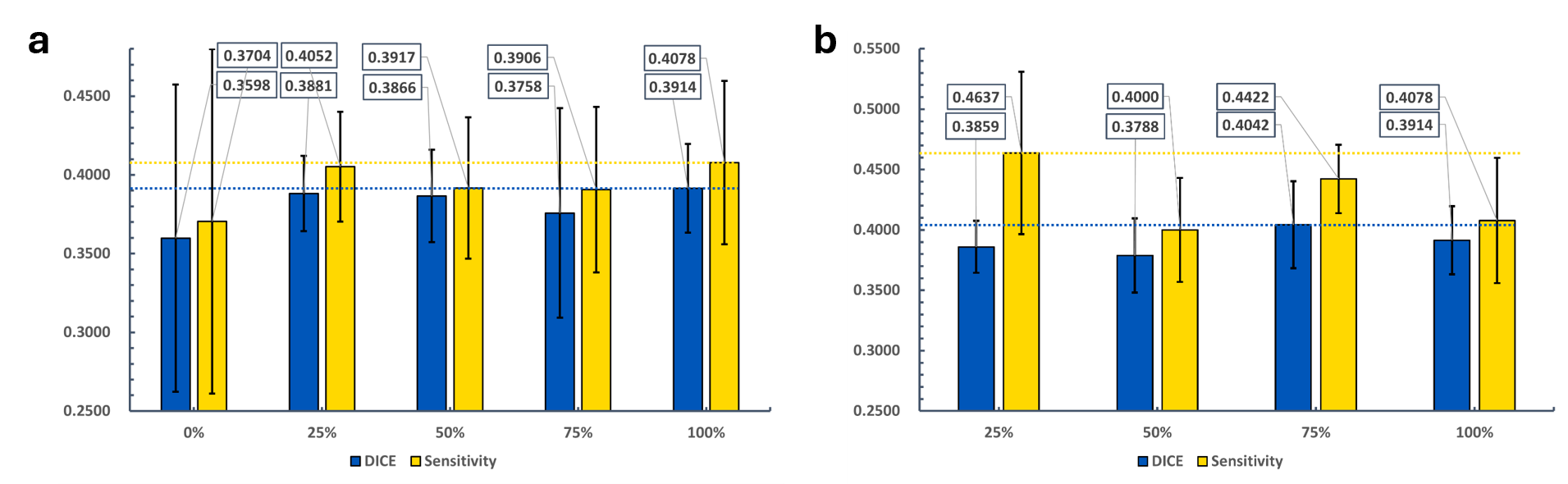}
\caption{Downstream lung nodule segmentation performance in different ratios of synthetic data and various data scarcity scenarios. \textbf{a}, Comparison of using different ratios of synthetic data for training a lung nodule segmentation model. \textbf{b}, Comparison of the synthetic data quality when using different ratios of real data for training the proposed method. Dashed lines indicate the best metrics in this evaluation.}
\label{fig5} 
\end{figure*}

\subsubsection{Lung nodule segmentation evaluation}
To further evaluate synthetic data quality, we design a challenging lung nodule segmentation evaluation to verify the synthetic data performance in a practical downstream task. Lung nodule segmentation traditionally followed a two-step approach: locating the rough nodule position, then performing detailed segmentation. Recently, this task has been simplified to focus solely on the second step, evaluating models using nodule-centered 3D CT image crops. State-of-the-art models have achieved Dice scores exceeding 0.9 in this simplified task \cite{dong2020multi,xiao2020segmentation,sun20203d}. To increase the challenge and more comprehensively assess segmentation models and synthetic image quality, we propose an end-to-end evaluation using complete thoracic CT images as input. This approach aims to push the performance boundaries of segmentation models and provide a more thorough test of fully synthetic thoracic CT image quality. 

In specific, we first sample 5-fold synthetic images according to the validation cohorts. For the IUCPQ dataset, we generate 800 samples per fold (4000 samples in total), and for the LIDC-IDRI dataset, 366 samples per fold (1830 samples in total). These synthetic images are then mixed with real samples from the discovery (upstream) cohort to form the discovery (downstream) cohort. We use 183, 400, and 800 synthetic samples in this cohort to test the lung nodule segmentation performance under different conditions of synthetic data usage. Next, we use the discovery (downstream) cohort to train an Attention U-Net \cite{oktay2018attention}, which is later tested on the test cohort. We also include three recently published competitors (MAV \cite{zhu2023make}, GEM-3D \cite{zhu2024generative}, and Med-DDPM \cite{dorjsembe2024conditional}) in this evaluation and compare the segmentation performance among the Attention U-Nets trained with synthetic data from these competitors and our proposed Lung-DDPM. The experimental results of the lung nodule segmentation evaluation are reported in Table \ref{table4}.

From Table \ref{table4}, we observe that the Dice scores decrease 11.3\%, 16.9\% and 1.7\% when incorporating a small number of synthetic data (183 samples) from MAV \cite{zhu2023make}, GEM-3D \cite{zhu2024generative}, and Med-DDPM \cite{dorjsembe2024conditional}) compared to the baseline (Dice: 0.3598, Sensitivity: 0.3704 and Specificity: 0.9999), respectively. Only Lung-DDPM achieves performance improvements in both Dice score (0.3908) and sensitivity (0.4393) under these conditions. When mixing 400 synthetic samples with 800 real samples to form the discovery cohort (downstream), Lung-DDPM also demonstrates superior performance in Dice scores (0.3866) and maintains acceptable sensitivity (0.3917) and specificity (0.9999). When a considerable number of synthetic data (800 samples) are included, only Lung-DDPM achieves noticeable improvements in both Dice scores (0.3914) and sensitivity (0.4078). These experimental results highlight that Lung-DDPM's synthetic samples can introduce more stable and appreciable lung nodule segmentation performance improvements compared to the other three SOTA generative models. This also provides additional evidence that samples generated by the proposed method have superior diversity and fidelity.

Moreover, Table \ref{table4} shows that adding real samples does not always lead to improved performance. For instance, incorporating 183 real samples from the external LIDC-IDRI dataset slightly decreases the Dice score from 0.3598 to 0.3571, indicating potential domain mismatch between the internal and external data. In contrast, adding 183 synthetic samples from Lung-DDPM yields a significantly higher Dice score of 0.3908 and improves sensitivity to 0.4393. Although adding 800 real samples from the internal IUCPQ validation cohort leads to the highest Dice score (0.4178), Lung-DDPM synthetic data still achieves comparable performance (Dice: 0.3914), offers close sensitivity (0.4078 vs. 0.4377), and preserves near-perfect specificity. These findings highlight that high-quality synthetic data from Lung-DDPM can outperform or rival real data augmentation in multiple scenarios, reinforcing its value for training segmentation models in data-scarce or privacy-constrained clinical environments.

\begin{table}[htbp] 
 \centering
 \caption{Ablation study of various proposed model structures and conditions. (The best evaluation score is marked in bold. $\uparrow$ means higher number is better, and $\downarrow$ indicates lower number is better. LS indicates lung semantic layouts.)} 
 \begin{tabular}{cccc} 
  \toprule 
  Experiments & FID ($\downarrow$) & MMD ($\downarrow$) & Dice (\%, $\uparrow$) \\ 
  \midrule 
  Ours & 0.0047±0.0080  & \textbf{0.0070±0.0002}
 & \textbf{0.3914±0.0282} \\
 w/o AAS & 0.0600±0.0040 & 0.6617±0.0260 & 0.3869±0.0450 \\
 w/o LS & \textbf{0.0013±0.0000} & 0.0107±0.0000 & 0.3827±0.0386 \\
 w/o AAS, LS & 0.0399±0.0015 & 0.4242±0.0167 & 0.3471±0.0851 \\
  \bottomrule 
 \end{tabular} 
\label{table5}
\end{table}

\subsubsection{Ablation studies}
To discuss the efficiency of the proposed AAS process and justify an optimal condition for guiding thoracic CT image synthesis, we conduct ablation studies following the experimental settings of the image quality evaluation and lung nodule segmentation evaluation. Experimental results, acquired using the IUCPQ dataset, are shown in Table \ref{table5}. In these studies, we investigate the performance contributions from the AAS process and the additional lung semantic layouts.

We observe that performance in both image quality and lung nodule segmentation evaluation decreases when we remove the AAS process, highlighting its considerable contribution (FID: $0.0047\rightarrow 0.0600$, MMD: $0.0070\rightarrow 0.6617$, Dice: $0.3914\rightarrow 0.3869$). When using only lung nodule semantic layouts as the condition, we observe an increased FID score ($0.0047\rightarrow 0.0013$). Without LS, the model relies solely on the lung nodule layout and the reference case during sampling. This reduces variation in the lung structure and leads to generation outcomes that more closely resemble the reference domain in global image statistics, which FID is sensitive to. Nonetheless, the absence of LS diminishes the model’s ability to control lung morphology explicitly, leading to poorer performance in downstream nodule segmentation, as evidenced by the reduced Dice Coefficient and sensitivity. However, using only lung nodule semantic layouts negatively impacts the Dice score in the downstream task (Dice: $0.3914\rightarrow 0.3827$). If we remove both the AAS process and use only lung nodule semantic layouts, the model performance decrease dramatically (FID: $0.0047\rightarrow 0.0399$, MMD: $0.0070\rightarrow 0.4242$, Dice: $0.3914\rightarrow 0.3471$).

Therefore, these ablation studies highlight that both the proposed AAS process and the proposed condition settings effectively contribute to synthetic image quality and downstream task performance.

\subsubsection{Finding the optimal ratio of synthetic data}
In this subsection, we discuss the optimal ratio of synthetic data in the training set for a lung nodule segmentation model. We evaluate the performance of a lung nodule segmentation model trained with 0\%, 25\%, 50\%, 75\%, and 100\% extra synthetic data from the IUCPQ dataset, as shown in Figure \ref{fig5}a. Compared to using only real data, incorporating extra synthetic data definitely introduces performance and stability improvements. We also observe that using 100\% extra synthetic data yields the best Dice score and sensitivity for lung nodule segmentation. However, using 25\% extra synthetic data achieves slightly reduced segmentation performance while offering better stability. Our findings suggest that 25\% synthetic data represents a sweet spot, offering a good trade-off between performance and computational complexity.

\begin{figure}[!ht]
\centering
\includegraphics[width=8cm]{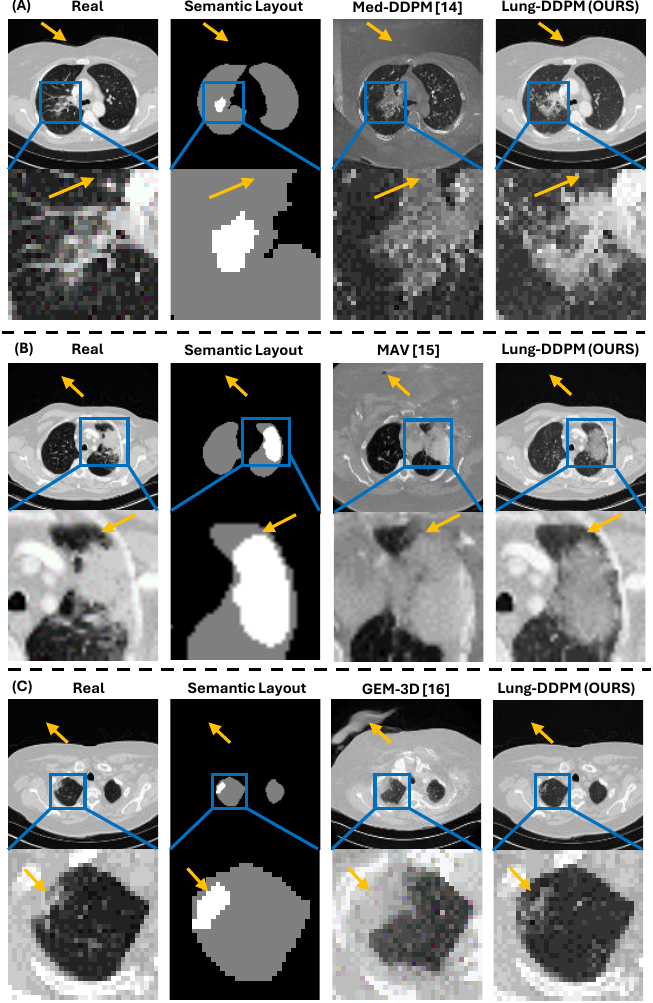}
\caption{Demonstration of how the proposed method addresses the anatomical imprecision issue, compared with three competitors. The columns represent: (1) real samples, (2) corresponding semantic layout (white: lung nodule, gray: lung, black: background), (3) synthetic samples generated by competitors, and (4) synthetic samples generated by the proposed method. For each sub-figure, the top row shows the original image, while the bottom row displays zoomed-in images. Yellow arrows indicate points of interest.}
\label{fig6} 
\end{figure}

\subsubsection{Performance in data scarcity scenarios}
The motivation for proposing Lung-DDPM is to improve lung nodule segmentation performance in data-scarce settings. To evaluate this, we reduced the size of the discovery (upstream) cohort to 25\%, 50\%, and 75\% to train three different Lung-DDPM generative models. We then used these pretrained models to generate synthetic samples according to the IUCPQ validation cohort. These synthetic samples, combined with real data, were used to train and evaluate three different lung nodule segmentation models. The results are reported in Figure \ref{fig5}b. 

Interestingly, segmentation models trained with synthetic data generated by a Lung-DDPM model trained on only 25\% of the real discovery cohort often achieve comparable or even superior performance to those using synthetic data generated from a model trained on the full 100\% discovery cohort. This counterintuitive result can be explained by considering the distribution of cases in the real dataset. The full 800-sample discovery cohort includes many small, low-contrast, ambiguous, or inconsistently labeled nodules that are inherently difficult to model and annotate accurately. Training the generative model on the full dataset exposes it to this variability, which can introduce artifacts or noise into the learned distribution and degrade the quality of generated samples. In contrast, the reduced 25\% subset may overrepresent clearer, larger, and more consistently labeled nodules that provide stronger, less ambiguous learning signals. This allows the generative model to learn a cleaner, more canonical distribution of nodule appearances. Consequently, synthetic samples generated from this model exhibit higher anatomical fidelity and more consistent structure, leading to better segmentation performance when used for augmentation. This finding highlights that, in data-scarce medical imaging scenarios where annotation quality varies significantly, carefully selecting high-quality training data for generative models can be more valuable than simply maximizing dataset size, and that high-fidelity synthetic data can effectively supplement real data to train robust segmentation models.

\subsection{Qualitative results}
We focus on presenting qualitative results in this section to intuitively demonstrate the proposed method's performance in addressing the following issues: I. Anatomical imprecision, II. Imperfect blending, and III. Inconsistency, while comparing it with the results from other state-of-the-art (SOTA) competitors.

In Figure \ref{fig6}, we demonstrate the existing anatomical imprecision issue using synthetic samples generated by Med-DDPM \cite{dorjsembe2024conditional}, MAV \cite{zhu2023make}, and GEM-3D \cite{zhu2024generative}, while showcasing the proposed method's performance in addressing this issue. In Figure \ref{fig6} (A), we observe that Med-DDPM generates unnecessary anatomical structures outside the human body. From the zoomed-in image, we also notice that Med-DDPM did not adhere to the semantic layout and generated an extra lung nodule structure. Leveraging the proposed AAS process, our method closely follows the real samples in the extra-pulmonary areas and generates lung nodules according to the provided semantic layout. In Figure \ref{fig6} (B), MAV fails to generate an anatomically reasonable extra-pulmonary area for the CT image, while our method synthesizes the correct extra-pulmonary structure by referencing real samples. Additionally, MAV generates lung nodule textures in areas that should be normal lung lobes according to the semantic layout. In contrast, our method reflects the semantic layout more accurately in lung nodule generation. In Figure \ref{fig6} (C), GEM-3D generates abnormal structures beyond the human body, while our method maintains a clean and anatomically reasonable target area. Regarding lung nodule generation, GEM-3D produces a larger lung nodule texture and overlooks the narrow lung lobe area between the lung nodule and thoracic tissues. However, our method preserves an anatomically reasonable structure while accurately generating the narrow lung lobe area. In summary, Lung-DDPM effectively addresses the anatomical imprecision issue that is prevalent among SOTA generative models and achieves improved performance in following semantic layouts by leveraging the proposed AAS process and the adjusted 3D U-Net structure.

\begin{figure}[!ht]
\centering
\includegraphics[width=8cm]{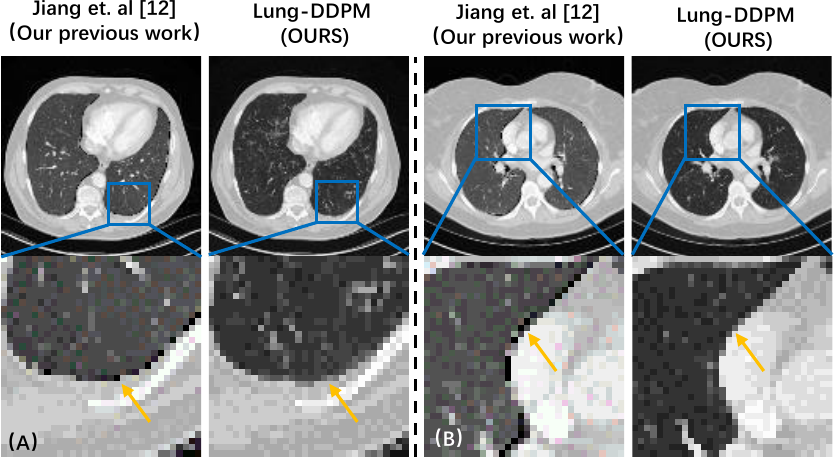}
\caption{Demonstration of how the proposed method solves the imperfect blending issue. In sug-figures (A) and (B), the first row shows samples generated by our previous method \cite{jiang2020covid} and the proposed method, respectively. The second row shows the zoom-in images. Yellow arrows indicate interesting points.}
\label{fig7} 
\end{figure}

In Figure \ref{fig7}, we demonstrate how the proposed method addresses the imperfect blending issue present in our previous work \cite{jiang2020covid}. Our previous method simply blended the synthetic lung image and the original extra-pulmonary images, which inevitably produced margins between these two areas because the generative model could not fully reflect the semantic layout, especially at the boundary between different labels. To overcome this limitation, instead of blending the synthetic lung image and the original extra-pulmonary images after the generation process, we propose using the AAS process to dynamically blend these two images during the sampling process. Experimental results in Figure \ref{fig7} indicate that the proposed method effectively avoids producing margins between synthetic and original images and generates synthetic CT images with higher quality.

\begin{figure*}[!ht]
\centering
\includegraphics[width=16cm]{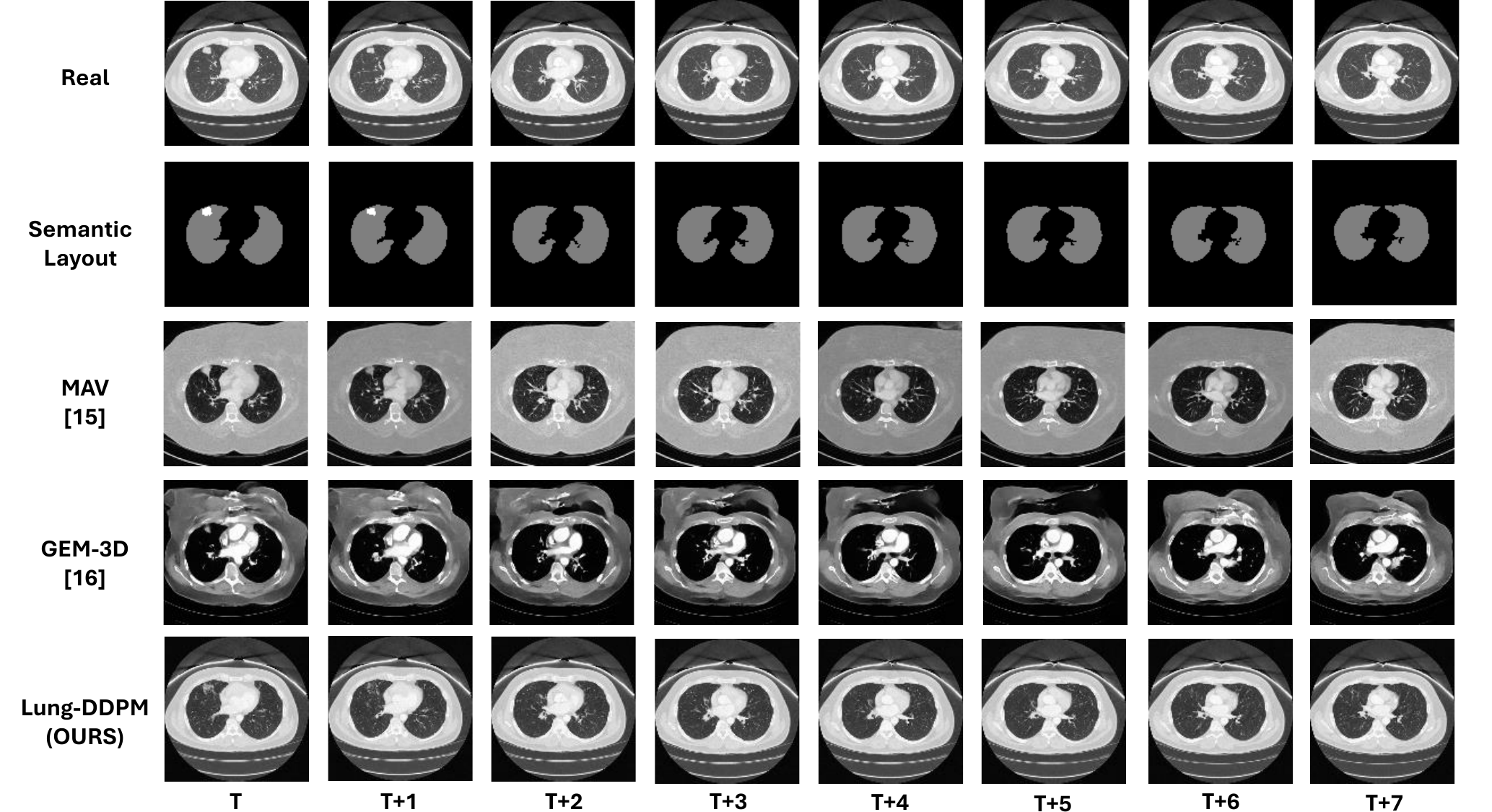}
\caption{Demonstration of how the proposed method solves the inconsistency issue. In the first and second rows, we present the real CT images and the corresponding semantic layouts (white: lung nodule, gray: lung, black: background). From the third to the fifth row, we demonstrate the synthetic samples generated by MAV \cite{zhu2023make}, GEM-3D \cite{zhu2024generative}, and the proposed method, respectively. In each column, we display a slice from a CT volume or a semantic layout volume that contains seven continuous slices.}
\label{fig8} 
\end{figure*}

In Figure \ref{fig8}, we illustrate how the proposed method solves the inconsistency issue that typically arises when applying partial generation methods such as MAV \cite{zhu2023make} and GEM-3D \cite{zhu2024generative}. Rather than performing end-to-end generation, these approaches generate partial CT images and then construct a complete image at the end of the generation process. This design choice makes the model lighter and enables the generation of CT images with higher resolution and more slices; however, it also introduces inconsistency when combining partially generated images. In Figure \ref{fig8},  we observe that MAV tends to produce synthetic samples with inconsistent contrast values. It is evident that slices at positions T, T+2, T+3, and T+7  are brighter than the others. This contrast inconsistency causes the synthetic CT image to "blink," significantly impacting image quality. In the case of GEM-3D, it generates structures with anatomical errors that abruptly change shape between slices T+5 and T+6. The sudden appearance or disappearance of an anatomical structure disrupts the consistency of the CT volume and negatively affects image quality. In contrast, the proposed method maintains strong consistency across CT slices from T to T+7 and better reflects the semantic layout compared to the other two competitors. To sum up, Lung-DDPM can generate synthetic CT images with stable consistency in both contrast and anatomical structures, which highlights its capabilities in thoracic CT image generation with better quality.

\section{Discussion}
\label{sec5}
Synthetic samples have been proven to alleviate data scarcity in training deep learning models for medical image analysis \cite{frid2018gan,zhang2023minimalgan,sherwani2024systematic}. However, using generative models to synthesize realistic and diverse thoracic CT images has long been a challenging task due to the complex and detailed anatomical features in the thoracic cavity. To accurately reflect the anatomical structures in synthetic CT scans, researchers have recently started incorporating semantic layouts as conditions in thoracic CT image synthesis. However, this approach has introduced new challenges \cite{dorjsembe2024conditional,zhu2023make,zhu2024generative}. The thoracic cavity is a complicated structure with multiple organs, making the annotation of semantic layouts costly and time-consuming cite{ma2021toward}. Consequently, researchers often have access to CT images with limited annotations. This difficulty in acquiring detailed semantic layouts directly contributes to the anatomical imprecision issue of synthetic samples and further hampers the development of generative models for thoracic CT image synthesis. Our findings underscore a practical solution to the anatomical imprecision issue in thoracic CT image synthesis. To avoid the expensive annotation process, we propose using only lung and lung nodule annotations. The former can be obtained reliably using automatic algorithms, while the latter is the only component requiring manual annotation. The proposed AAS process dynamically blends extra-pulmonary reference images with synthetic images generated according to these annotations to maintain precise anatomical structures while simultaneously introducing diversity into synthetic samples. Compared to other state-of-the-art (SOTA) generative models, the proposed method demonstrates a strong capability in addressing the anatomical imprecision issue (Figure \ref{fig6}). Additionally, its ability to generate anatomically precise CT images allows the proposed method to achieve better image quality (Table \ref{table3}) and significantly contribute to downstream lung nodule segmentation tasks (Table \ref{table4}).

To address the anatomical imprecision issue, our previous work \cite{jiang2020covid} attempted to blend extra-pulmonary reference images with synthetic lung images using a simple image fusion approach. While this method alleviated some concerns, it introduced a new problem: imperfect blending. Since generative models typically cannot fully reflect the semantic layouts, this limitation results in margins at the edges of synthetic samples, significantly impacting image quality. Although applying additional image fusion techniques \cite{perez2023poisson,amolins2007wavelet} may help mitigate imperfect blending, these approaches often increase computational complexity. In contrast, Lung-DDPM employs the AAS process to blend synthetic and reference images dynamically during the sampling process, effectively resolving the imperfect blending problem without incurring extra computational costs. In Figure \ref{fig7}, the comparison between synthetic samples generated by Lung-DDPM and our previous works illustrates that Lung-DDPM achieves more natural blending results and higher image quality.

Unlike natural images, medical images are typically represented in a three-dimensional format rather than two-dimensional. The additional Z-axis is essential for stacking multiple 2D slices to form a coherent 3D representation of the scanned anatomy. Therefore, it is crucial for generative models used in thoracic CT image synthesis to maintain consistency along the Z-axis. However, the most recent SOTA methods \cite{zhu2023make, zhu2024generative} encounter inconsistency issues due to their partial sampling strategy. Specifically, these approaches divide a 3D thoracic CT image into several parts along the Z-axis, sampling each part individually and then assembling them based on their overlaps. While this partial sampling strategy allows for lighter network designs, it sacrifices consistency. In contrast, the proposed AAS strategy enables Lung-DDPM to generate a complete thoracic CT image in an end-to-end manner, ensuring consistency along the Z-axis (Figure \ref{fig8}). This approach not only improves image quality (Table \ref{table3}) but also enhances performance in downstream lung nodule segmentation tasks (Table \ref{table4}).

How to effectively utilize synthetic samples is a critical topic worth discussing. Existing studies have concluded that an appropriate amount of synthetic samples can be beneficial to downstream task performance \cite{garcea2023data}. However, when narrowing our focus to lung nodule segmentation, the answer becomes ambiguous due to a lack of evidence. In Figure \ref{fig5}a, we confirm that a mixed training set for the lung nodule segmentation model, consisting of 75\% real samples and 25\% synthetic samples, is the most reliable and stable configuration.

Additionally, data scarcity can vary depending on actual conditions. Under extreme data scarcity scenarios, deep generative models face more challenges due to the limited number and diversity of training data \cite{alzubaidi2023survey}. As demonstrated in Figure \ref{fig5}b, reducing the training data amount to 25\% results in increased instability in the downstream task's sensitivity (indicated by larger error bars), despite achieving the highest score. Conversely, as we gradually increase the training data amount, sensitivity becomes more stable (indicated by smaller error bars), although it remains sub-optimal. The Dice score remains at a similar level across different amounts of training data. These experimental results highlight the importance of utilizing as many real samples as possible. Notably, Lung-DDPM can effectively handle challenging data scarcity scenarios, even with only 25\% of real samples remaining.

\section{Conclusion and future study}
\label{sec6}
In this paper, we introduced Lung-DDPM, a DDPM-based thoracic CT image synthesis method capable of generating 3D synthetic CT images with high diversity and fidelity. The proposed method incorporates two key components for conditional thoracic CT image synthesis: a novel anatomically aware sampling (AAS) process and an adjusted 3D U-Net denoiser structure. Through the proposed AAS process, Lung-DDPM can generate realistic thoracic CT images with precise anatomical structures and acceptable diversity. Additionally, Lung-DDPM effectively addresses three prevalent issues in SOTA generative models when performing the conditional thoracic CT image synthesis task: anatomical imprecision, imperfect blending, and inconsistency. The experimental results indicate that the proposed method outperforms other SOTA generative models in both image quality and downstream lung nodule segmentation evaluations. This highlights its strong capability to generate high-quality synthetic CT images that can be beneficial in various downstream diagnostic tasks. Although Lung-DDPM provides high-quality volumetric synthesis and spatial consistency, we recognize that computational efficiency and scalability are crucial for practical deployment. In future work, we plan to conduct a comprehensive study on the trade-offs between synthesis quality and computational demands, and to further optimize the Lung-DDPM framework for memory and runtime efficiency.

\section{Acknowledgments}
\label{sec7}
Venkata Manem holds a salary support award from the IVADO and was supported by the New Frontier Research - Rapid Response Fund. This work is funded by the IUCPQ Biobank and foundation, and the SynergiQC project/CQDM. 

\section*{References}

\bibliographystyle{IEEEtran}
\bibliography{IEEEtran}{}

\end{document}